%% file: ISMA2026.tex
\documentclass{ISMA_USD}

\usepackage{booktabs}
\usepackage{threeparttable}
\usepackage{xcolor}
\usepackage{gensymb}
\usepackage{colortbl}
\usepackage{textcomp}
\usepackage{csquotes}
\usepackage{placeins}
\usepackage{adjustbox}

\hypersetup{
  pdftitle={Robust and Explainable 3D Mode Shape Recognition Using Region-Aware Graph Neural Networks},
  pdfauthor={Tong Duy Son, Marc Brughmans, Andrey Hense, Kohta Sugiura, Sebastian Ciceo, Paolo di Carlo, Theo Geluk},
  pdfkeywords={Mode Shape Recognition, Graph Neural Networks, Explainable AI, Body-in-White, NVH, CAE}
}

\title{Robust and Explainable 3D Mode Shape Recognition \\ Using Region-Aware Graph Neural Networks}

\author[1]{Tong Duy Son}
\author[1]{Marc Brughmans}
\author[1,2]{Andrey Hense}
\author[3]{Kohta Sugiura}
\author[1]{Sebastian Ciceo}
\author[1]{Paolo di Carlo}
\author[1]{Theo Geluk}

\affil[1] {Siemens Digital Industries Software,\NewLineAffil
      Interleuvenlaan 68, 3001 Leuven, Belgium \NewLineAffil
e-mail: \textbf{son.tong@siemens.com}\NewAffil}

\affil[2] {KU Leuven, Campus Diepenbeek, Department of Mechanical Engineering \NewLineAffil
      Wetenschapspark 27, B-3590, Diepenbeek, Belgium \NewAffil}

\affil[3] {Siemens Digital Industries Software, \NewLineAffil
      3-1-9 Shin-Yokohama, Yokohama 222-0033, Japan \NewAffil}

\date{}

\begin{document}

\abstract{
  \\
Body-in-White (BiW) mode shape recognition is a fundamental task in automotive noise, vibration, and harshness (NVH) development, 
yet it remains dependent on manual visual inspection by experienced engineers. 
Existing approaches based on engineering heuristics, Modal Assurance Criterion (MAC), or geometry-dependent AI representations 
often exhibit limited robustness across different vehicle architectures, finite element (FE) meshes, and experimental measurement layouts, 
restricting their applicability in industrial development.
\\[6pt]
This paper presents a canonical engineering graph representation and region-aware graph learning framework for robust and explainable 
3D mode shape recognition. Rather than learning directly from vehicle-specific FE meshes, heterogeneous FE models and experimental 
measurements are transformed into a common graph whose nodes represent semantically meaningful structural regions connected through 
engineering-informed relationships. Geometry-independent regional descriptors are combined with graph attention learning and region-aware 
pooling to capture structural interactions while preserving engineering semantics and enabling physically interpretable predictions. 
The resulting representation decouples engineering knowledge from numerical discretization, 
allowing learning to transfer across different vehicle programs without requiring identical mesh topology or sensor configurations.
\\[6pt]
The proposed framework is validated using FE and experimental datasets from four vehicle programs under severe label scarcity. 
Experimental results demonstrate high classification accuracy, cross-vehicle transferability, 
and physically meaningful explanations by directly relating model predictions to engineering-defined structural regions used in NVH analysis. 
Beyond mode shape recognition, the proposed Canonical Engineering Graph Representation establishes a reusable engineering abstraction 
that can support trustworthy and transferable AI across heterogeneous simulation and experimental workflows.
}

\maketitle

\input{introduction}
\input{background}
\input{framework}
\input{datageneration}

\input{experiment}
\input{conclusion}

\bibliography{main}

\end{document}

%% file: introduction.tex

\section{Introduction}

Automotive product development relies heavily on simulation-driven engineering,
where computer-aided engineering (CAE) and computational fluid dynamics (CFD)
are used to evaluate structural, vibro-acoustic, and aerodynamic performance
before physical prototypes are finalized. In structural development, finite
element (FE) models are widely used to analyze body stiffness, modal behaviour,
and noise, vibration, and harshness (NVH), while aerodynamic simulations are
used to predict drag, pressure distribution, wall shear stress, and flow
behaviour around the vehicle \cite{ewins2000modal, son2026generalizable}. These analyses play a
central role in engineering decisions related to body design,
structural refinement, aerodynamic efficiency, and overall vehicle
performance.

Modern vehicle development generates a large volume of digital engineering
artifacts throughout the design, simulation, testing, and validation process.
These include 3D CAD models, Body-in-White (BiW) and trimmed-body
representations, finite element models, CFD meshes, and experimental testing
measurements. In practice, these engineering assets are distributed across
different software tools, vehicle programs, simulation variants, sensor
layouts, and development teams. Consequently, valuable engineering knowledge
often remains fragmented across heterogeneous engineering representations,
making it difficult to reuse historical engineering data and develop AI
methods that operate consistently throughout the engineering workflow.

\begin{figure}[t]
  \centering
  \includegraphics[width=\linewidth]{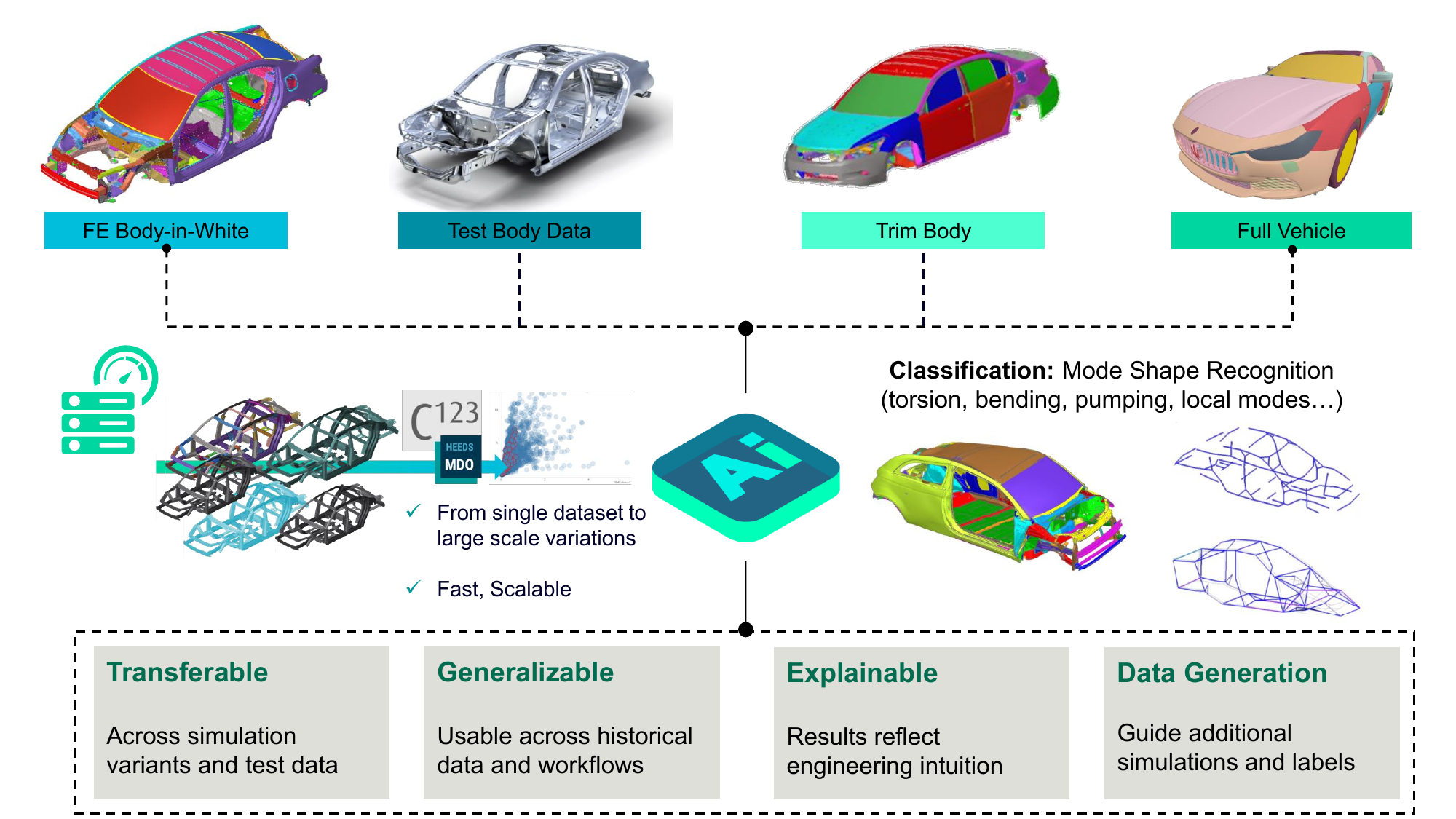}
  \caption{Overview of the proposed Canonical Engineering Graph Representation 
  and region-aware graph learning framework for explainable and
  transferable Body-in-White mode shape classification.}
  \label{fig:main_diagram}
\end{figure}

Recent advances in artificial intelligence (AI) have created new opportunities
to accelerate engineering workflows by reducing repetitive expert effort and
supporting engineering decision making. In structural dynamics, AI has been
applied to structural interpretation, mode assessment, cross-variant
comparison, and data-driven modelling of NVH behaviours
\cite{tohmuang2025modegcn,Millan10112025}. In particular, automatic mode shape
classification has attracted increasing interest because engineers routinely
inspect hundreds of vibration modes throughout vehicle development. Despite
decades of progress in simulation technology, manual interpretation of modal
deformation patterns remains time consuming, subjective, and difficult to
scale across multiple vehicle programs and design iterations.

Graph neural networks (GNNs) have recently emerged as a promising learning
paradigm for 3D engineering AI because they preserve structural
relationships while operating on irregular engineering data. Compared with
image- or voxel-based representations, graph representations explicitly model
structural connectivity and physical interactions, making them particularly
well suited to finite element models and other 3D engineering
simulation models. Existing graph-based methods have demonstrated promising performance 
across structural mechanics, surrogate modelling, and physics-based simulation
\cite{pfaff2020learning}. Nevertheless, most existing approaches remain
closely coupled to geometry-dependent representations, making their
performance sensitive to mesh discretization, node layouts, and sensor
configurations. As a result, transferring learned models across different
vehicle architectures or between simulation and experimental data remains a
challenging problem.

From an engineering perspective, this limitation reflects a more fundamental
challenge. Engineers rarely interpret structural behaviour at the level of
individual finite element nodes. Instead, modal assessment is naturally
performed using persistent structural entities such as roof rails, pillars,
side sills, floor structures, and cross members. While numerical
discretizations may vary substantially across vehicle programs, these
engineering concepts remain significantly consistent. This observation suggests
that transferable engineering AI should learn from engineering semantics
rather than geometry-dependent numerical representations.

For practical deployment in industrial NVH workflows, four challenges remain
particularly important:

\begin{enumerate}
\setlength{\itemsep}{0pt}

\item \textbf{Transferability:} AI models should generalize across different
vehicle programs, historical datasets, finite element discretizations, and
experimental sensor layouts without requiring extensive vehicle-specific
retraining.

\item \textbf{Limited labelled data:} obtaining expert-reviewed mode labels is
expensive, making it difficult to construct large annotated datasets.

\item \textbf{Explainability:} predictions should be physically interpretable
and consistent with the engineering reasoning used during modal assessment,
allowing engineers to understand the structural mechanisms behind each
prediction.

\item \textbf{Industrial practicality:} AI methods should integrate
throughout the engineering development workflow by enabling consistent
reasoning across heterogeneous engineering representations, including
simulation models, experimental measurements, test data, and successive
vehicle programs, while remaining robust under limited annotations.

\end{enumerate}

Despite encouraging progress, current graph learning pipelines remain largely
vehicle specific because they learn directly from geometry-dependent finite
element meshes rather than engineering concepts. Consequently, model
predictions often provide limited physical insight and are difficult to reuse
across different vehicle programs, simulation variants, or experimental
measurements. More importantly, relatively little attention has been devoted
to developing reusable engineering representations that preserve engineering 
semantics across heterogeneous datasets.

To address these challenges, this paper proposes a region-aware graph learning
framework built upon a Canonical Engineering Graph Representation for
robust and explainable BiW mode shape recognition. Rather than
learning directly from geometry-dependent finite element meshes, the proposed
approach first constructs a graph representation that transforms heterogeneous engineering
models into a common semantic graph composed of engineering-defined structural
regions connected through physically meaningful relationships. Regional
displacement responses are aggregated into geometry-independent node features,
enabling structurally consistent representations across different vehicle
architectures and sensor layouts. Graph attention learning captures structural
interactions between these regions, while a region-aware pooling strategy
incorporates engineering-informed structural descriptors to improve
discrimination between physically similar mode families. By separating
engineering semantics from numerical discretization, the proposed framework
enables transferable graph learning while naturally supporting
engineering-oriented interpretation of model predictions.

The proposed framework is validated using BiW simulation and experimental
datasets from multiple vehicle programs. The results demonstrate robust
classification performance despite severe label scarcity while maintaining
strong transferability across different vehicle architectures, mesh
discretizations, and sensor layouts. Furthermore, the semantic graph
representation enables physically meaningful explanations by relating model
predictions directly to structural regions commonly used during engineering
review. Although this work focuses on hierarchical mode shape recognition,
the proposed canonical engineering graph provides a reusable engineering
representation that can support trustworthy and scalable Engineering AI across
automotive development workflows.

The main contributions of this work are summarized as follows:

\begin{enumerate}
\setlength{\itemsep}{0pt}

\item We propose a canonical engineering graph representation that transforms heterogeneous 
BiW models into a common semantic space independent of mesh topology and sensor configuration, 
enabling robust learning across different vehicle
architectures, finite element discretizations, and experimental measurement
configurations.

\item We develop a region-aware graph learning framework that combines graph attention learning
with engineering-informed regional descriptors for explainable hierarchical
mode shape classification.

\item We demonstrate robust cross-vehicle transfer using both finite element
and experimental datasets under severe label scarcity while providing
physically interpretable predictions that can be directly related to
meaningful structural regions used in engineering practice.

\end{enumerate}

The remainder of this paper is organized as follows. Section 2 reviews related
work on CAE mode shape classification and graph learning. Section 3 presents
the proposed framework. Section 4 describes the dataset generation.
Section 5 presents the experimental results. Finally,
Section 6 concludes the paper and discusses future works.

%% file: background.tex

\section{Related Work}

This section reviews previous research most relevant to automatic mode shape
recognition and graph representations for 3D engineering structures. Rather than
providing a broad survey of engineering AI, the discussion focuses on
developments that directly motivate the proposed Canonical Engineering Graph
Representation and region-aware graph learning framework for explainable Body-
in-White (BiW) mode shape classification.

\subsection{Automated Mode Shape Recognition}

Automatic recognition of structural vibration mode shapes has long been an
important objective in automotive CAE. During vehicle development, engineers
routinely inspect finite element (FE) modal results to identify characteristic
deformation mechanisms, including torsional, bending, pumping, and local
structural modes. These modal characteristics provide essential insight into
body stiffness, structural dynamics, and noise, vibration, and harshness (NVH)
performance while supporting design optimisation across successive vehicle
programs \cite{ewins2000modal,Millan10112025}.

Despite decades of development in simulation technology, mode classification
remains heavily dependent on manual engineering interpretation. Engineers
visually compare modal deformation patterns and assign engineering labels based
on structural experience. Although reliable, this process is time consuming,
subjective, and increasingly difficult to scale when hundreds of vibration
modes must be reviewed across multiple design iterations and vehicle variants.

Early efforts to automate this task relied primarily on engineering heuristics,
Modal Assurance Criterion (MAC), modal parameter estimation, and manually
designed structural descriptors extracted from simulation results
\cite{ewins2000modal}. Classical machine learning methods subsequently improved
automation by learning from handcrafted features, but their performance
remained closely tied to vehicle-specific preprocessing and feature
engineering. More recently, graph-based learning methods have emerged as a
promising alternative for structure-aware mode shape recognition, with
\cite{tohmuang2025modegcn} demonstrating that graph convolutional networks can
successfully classify structural mode shapes directly from engineering
representations, highlighting the potential of graph learning for industrial
NVH applications.

However, several limitations remain. Most existing graph-based approaches are
coupled to individual vehicle models, limiting transfer across different vehicle
architectures and mesh discretisations. Furthermore, classification accuracy
alone provides limited engineering value without interpretable predictions, and
comparatively little attention has been given to developing reusable
representations that enable systematic knowledge transfer across vehicle
programs.

\subsection{Graph Representations and Learning for Engineering Structures}

Graph neural networks (GNNs) have emerged as an effective learning paradigm for
engineering applications because they naturally represent irregular spatial
relationships while preserving structural connectivity. Compared with regular
grids or images, graph representations allow engineering entities to be
modelled together with their physical interactions, making them well suited to
finite element models, computational meshes, and other simulation-derived
engineering data.

Several alternative representations have been explored for three-dimensional
engineering problems. Voxel and grid-based methods are compatible with
convolutional neural networks but often require fine spatial discretisation to
preserve geometric fidelity, resulting in high computational and memory costs
for large engineering models. Point-based approaches, including PointNet and
PointNet++, avoid explicit meshing and have demonstrated strong performance on
geometric recognition tasks, but they do not explicitly model structural
connectivity or engineering relationships \cite{qi2017pointnet}. Graph
representations, by contrast, explicitly preserve adjacency, connectivity, and
physical interactions through node-edge relationships, making them
particularly suitable for structural mechanics and physics-based engineering
problems. Recent advances have further extended graph learning through
attention mechanisms, graph transformers, physics-informed learning, and
mesh-based surrogate modelling, demonstrating promising performance for
structural analysis, computational mechanics, and engineering simulation
\cite{pfaff2020learning,velickovic2018graph,park2024bmognn,taghizadeh2025multifidelity}.

The success of these methods has established graph learning as an important
computational framework for Engineering AI. However, their effectiveness
depends not only on the learning architecture but also on how engineering
systems are represented as graphs. Different graph construction strategies
encode different structural relationships, physical priors, and engineering
knowledge, which can significantly influence prediction performance,
generalisation capability, transferability, and interpretability.

Most existing approaches construct graphs directly from computational meshes,
finite element discretisations, or geometric neighbourhoods. Consequently, the
graph topology largely reflects numerical discretisation rather than the
engineering concepts used during product development. Variations in mesh
topology, node numbering, element density, or measurement layouts therefore
often require adapting the graph representation, limiting knowledge reuse
across different vehicle programs, simulation variants, and experimental
measurements.

From an engineering perspective, however, structural behaviour is rarely
interpreted at the level of individual finite element nodes. Instead,
engineers reason using persistent structural entities such as roof rails,
pillars, side sills, floor structures, cross members, and longitudinal load
paths. These engineering concepts remain largely consistent across different
vehicle programs even when the underlying numerical discretisations differ
substantially. This observation suggests that transferable Engineering AI
should learn from engineering semantics rather than geometry-dependent
numerical representations.

Existing studies therefore demonstrate the effectiveness of graph learning for
engineering applications while leaving the design of reusable engineering
representations largely unexplored \cite{son2026generalizable}. This gap motivates the Canonical
Engineering Graph Representation proposed in this work, which represents
heterogeneous engineering models using persistent engineering regions and
physically meaningful structural relationships rather than vehicle-specific
numerical discretisations.

\subsection{Research Gap and Motivation}

The preceding literature review highlights three remaining challenges that
motivate the present work.

First, existing graph learning methods generally operate on
geometry-dependent representations that remain sensitive to mesh topology,
vehicle-specific discretisations, and sensor layouts. This limits their ability
to transfer across different vehicle programs without substantial retraining.
Second, although recent GNNs achieve encouraging classification performance,
their learned representations are typically difficult to interpret from an
engineering perspective. Attention maps or node importance scores alone do not
necessarily correspond to the structural entities that engineers use during
modal assessment and design review. Third, most current studies focus primarily
on improving predictive performance, while comparatively little attention has
been devoted to developing reusable engineering representations that preserve
engineering semantics across simulation models, experimental measurements, and
successive vehicle generations. As a result, historical engineering knowledge
cannot be readily reused and is often relearned independently for each vehicle
program instead of being accumulated within a common engineering
representation.

These observations motivate the region-aware graph framework proposed in this
paper. Rather than learning directly from geometry-dependent finite element
meshes, the proposed approach constructs a Canonical Engineering Graph
Representation whose nodes represent persistent engineering-defined structural
regions and whose edges encode physically meaningful structural relationships.
By separating engineering semantics from numerical discretisation, the proposed
representation provides a common foundation for transferable graph learning,
engineering-oriented interpretation, and knowledge reuse across heterogeneous
engineering datasets.

The central hypothesis of this work is that transferable learning across
heterogeneous engineering datasets is enabled not primarily by increasingly
sophisticated graph neural network architectures, but by learning on a
canonical engineering graph representation that preserves engineering semantics
while remaining independent of mesh discretisation.

%% file: framework.tex

\section{Proposed Region-Aware Graph Framework}

\begin{figure}
  \centering
  \includegraphics[width=\linewidth]{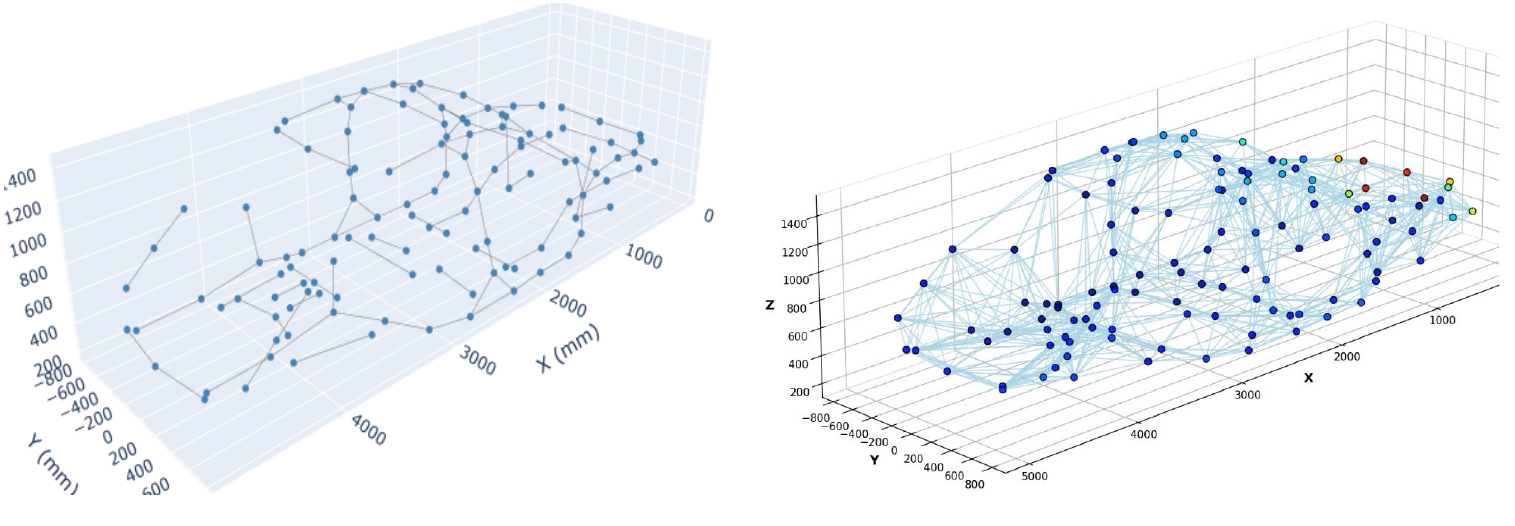}
  \caption{Graph construction from wireframe.}
  \label{fig:uc1_graph}
\end{figure}

The proposed framework introduces a \emph{Canonical Engineering Graph
Representation} that maps heterogeneous engineering models into a common
semantic graph whose nodes represent persistent structural regions and whose
edges encode physically meaningful relationships. By decoupling engineering
semantics from numerical discretisation, this representation provides a
transferable foundation for graph learning across different vehicle programs,
mesh topologies, and experimental configurations.

This design is motivated by an observation from industrial NVH development.
Although BiW models differ substantially across vehicle programs in mesh
resolution, modelling strategy, assembly configuration, and sensor layout,
engineers interpret vibration behaviour through persistent structural entities
such as roof rails, A/B/C-pillars, side sills, floor structures, cross
members, and rear body components. These entities remain largely consistent
across multiple vehicle programs even when the underlying numerical models
change significantly.

A representation is considered \emph{canonical} in this work if engineering
entities performing equivalent structural functions are mapped to the same
graph entities regardless of vehicle geometry, mesh topology, node numbering,
or experimental measurement configuration. Knowledge is therefore transferred
at the level of engineering concepts rather than numerical mesh entities.
Based on this philosophy, the framework is designed according to three
engineering objectives.

\begin{figure}
  \centering
  \adjustbox{frame=0.1pt, width=0.75\linewidth, height=7cm}{%
    \includegraphics{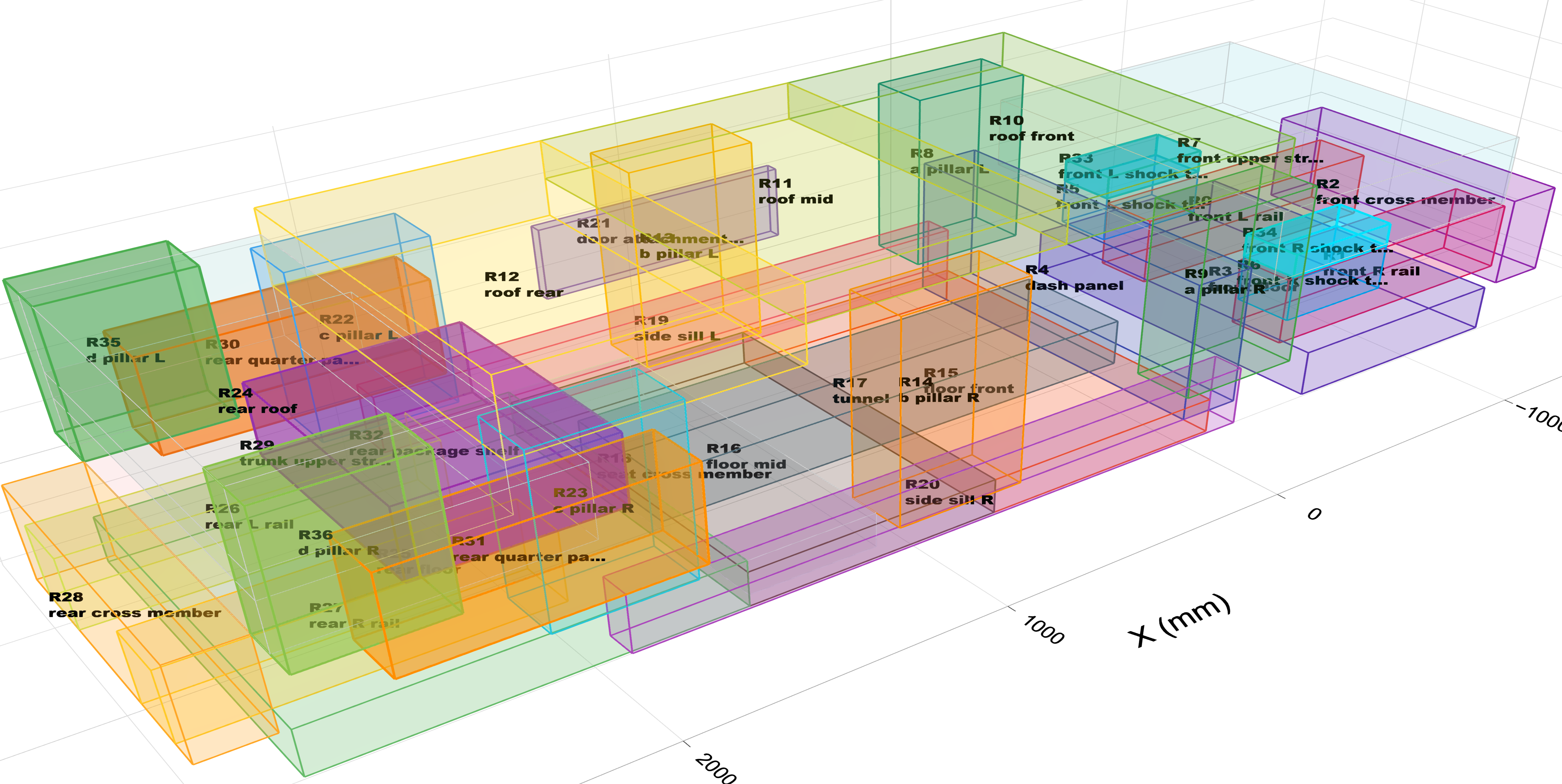}}\\[6pt]
  \caption{Canonical BiW regional decomposition used for
  engineering-aware feature fusion and aggregation.}
  \label{fig:uc1_regions}
\end{figure}

\textbf{Engineering semantics.}
Nodes and edges correspond to recognisable structural regions and physically
meaningful relationships rather than arbitrary finite element nodes or mesh
connectivity.

\textbf{Transferable representations.}
The framework remains applicable across different vehicle programs, finite
element discretisations, assembly variants, and experimental measurements
without requiring fundamentally different graph constructions.

\textbf{Engineering interpretability.}
Beyond accurate prediction, industrial deployment requires that engineers
understand which structural regions contribute to a classification and whether
these regions correspond to physically meaningful deformation mechanisms.

Guided by these objectives, the framework consists of four stages:
\begin{enumerate}
\setlength{\itemsep}{0pt}

\item \textbf{Canonical Engineering Graph Representation:}
heterogeneous engineering models are transformed into a common semantic graph
whose nodes represent engineering-defined structural regions and whose edges
encode engineering-informed structural relationships.

\item \textbf{Graph representation learning:}
a graph attention network propagates information between engineering regions
through message passing on the canonical graph, learning structural
interactions associated with each vibration mode.

\item \textbf{Engineering-aware prediction:}
learned graph embeddings are fused with engineering-informed regional
descriptors for hierarchical mode shape classification.

\item \textbf{Engineering interpretation and knowledge reuse:}
graph-level predictions are mapped back to engineering-defined structural
regions to support explainable model interpretation, cross-vehicle transfer,
and reuse of learned engineering knowledge.

\end{enumerate}

Figure~\ref{fig:main_diagram} summarises the complete workflow. The following
subsections describe the Canonical Engineering Graph Representation, the
region-aware graph learning framework, and the resulting engineering
properties for explainability and transferability.

\subsection{Canonical Engineering Graph Representation}

\begin{figure}
  \centering
  \adjustbox{frame=0.1pt, width=0.75\linewidth, height=7cm}{%
    \includegraphics{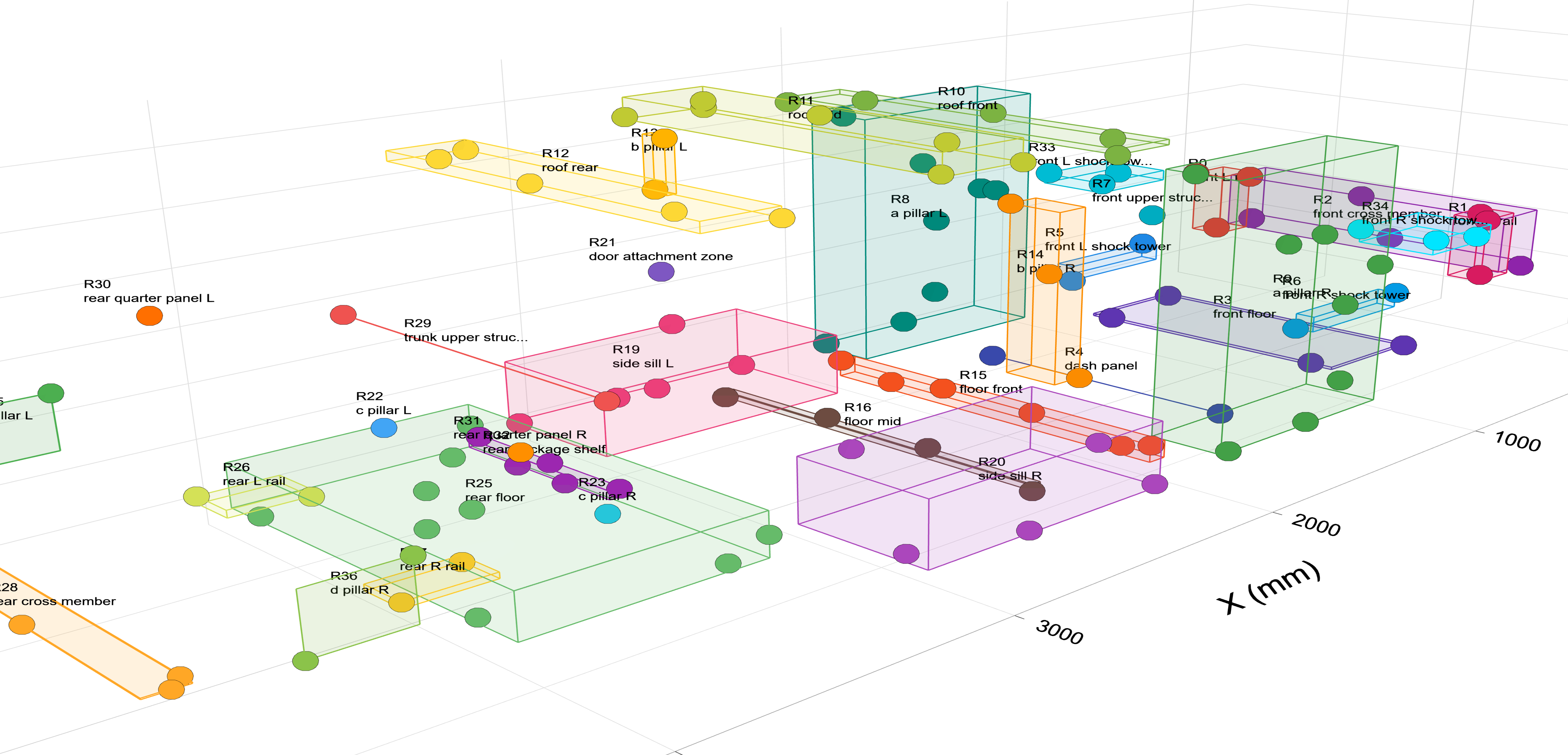}}
  \caption{Region-aware aggregation of a vehicle wireframe model with
  119 nodes onto the canonical engineering graph.}
  \label{fig:uc1_regions_vehicle}
\end{figure}

The Canonical Engineering Graph Representation is formalized as a mapping
from a heterogeneous engineering model to a semantic graph. Starting from a
finite element model or experimental wireframe, the structural skeleton of the
vehicle is extracted and partitioned into persistent BiW regions, including
front rails, side sills, roof rails, A/B/C-pillars, floor structures, cross
members, and rear body components, which correspond to the longitudinal load
paths, pillars, and lateral members described in the engineering review
process. Unlike finite element nodes, these engineering regions remain
consistent across vehicle architectures and modelling strategies, making them
suitable as transferable graph entities.

Each vehicle is then represented as the attributed graph

\[
G = \left(\mathcal{V},\, \mathcal{E},\, \mathbf{X},\, \mathbf{R}\right),
\]

where $\mathcal{V}$ denotes the set of engineering-defined structural regions
(rather than individual finite element nodes or measurement sensors),
$\mathcal{E}$ represents engineering-informed structural relationships between
regions, $\mathbf{X}$ contains regional node features, and $\mathbf{R}$ stores
edge attributes describing pairwise structural interactions. Each node
summarises the collective dynamic behaviour of one engineering region, enabling
finite element models with different mesh densities, topologies, and numbering
schemes to be represented using a common semantic graph structure.

Four categories of structural relationships are considered as graph edges:
structural adjacency between neighbouring body regions, left-right symmetry,
longitudinal load-path coupling, and vertical roof-floor coupling. These
relationships reflect how structural deformation propagates throughout the
vehicle body and provide stronger engineering priors than purely
distance-based graph construction.

For each engineering region, displacement responses from the original finite
element model are aggregated to compute geometry-independent regional
descriptors. The node feature vector comprises the following quantities: mean
displacement magnitude, RMS displacement magnitude, dominant displacement
direction, signed vertical response, and regional deformation energy. These
features characterise the modal behaviour of each engineering region
independently of the underlying mesh discretisation.

Edge attributes provide complementary relational information describing the
interaction between neighbouring structural regions, including edge type,
relative deformation energy, phase agreement, and structural coupling
characteristics. Together, node and edge attributes allow the graph to capture
both local structural behaviour and global deformation mechanisms while
preserving explicit engineering semantics.

The resulting representation possesses three fundamental properties. First,
it is \textbf{geometry independent}, allowing different finite element meshes
and experimental sensor layouts to be mapped onto a common semantic graph
structure. Second, it is \textbf{engineering interpretable}, since every graph
entity corresponds directly to recognisable structural members used during NVH
engineering review. Third, it is \textbf{knowledge transferable}, because
learning is performed on persistent engineering concepts rather than numerical
mesh entities. Once engineering models have been transformed into this common
representation, different graph learning algorithms can operate on the same
engineering abstraction across simulation models, experimental measurements,
and multiple vehicle programs.

\subsection{Region-Aware Graph Learning}

Given the Canonical Engineering Graph Representation defined above, the
objective of the learning framework is to infer structural mode categories from
regional dynamic responses while preserving engineering interpretability. The
goal is not to introduce a new graph neural network architecture, but to enable
graph learning on a representation whose nodes and edges already carry
engineering meaning.

Each mode shape is represented as one graph sample. Node attributes describe
regional dynamic responses, while edges encode engineering-defined structural
relationships. The graph is processed using a multi-layer graph attention
network (GAT), which propagates information between neighbouring engineering
regions through message passing.

For graph layer $l$, the hidden representation of node $i$ is updated as

\[
\mathbf{h}_i^{(l+1)}
=
\sigma
\!\left(
\sum_{j \in \mathcal{N}(i)\,\cup\,\{i\}}
\alpha_{ij}^{(l)}\,
\mathbf{W}^{(l)}
\mathbf{h}_j^{(l)}
\right),
\]

where $\mathbf{h}_i^{(l)}$ denotes the node embedding at layer $l$,
$\mathbf{W}^{(l)}$ is a learnable linear transformation matrix, and
$\sigma(\cdot)$ denotes a nonlinear activation function. The self-loop term
$j = i$ is included so that each region retains its own embedding during
aggregation. Attention coefficients $\alpha_{ij}^{(l)}$ are computed by
applying a shared attention function to the concatenated features of each node
pair, with the resulting scores normalized across all neighbours using a
softmax function. Attention allows the network to assign different importance
to neighbouring structural regions depending on the current deformation
pattern. After several graph attention layers, each node embedding captures
both local regional response and structural context within the BiW graph.

A conventional graph classifier would aggregate node embeddings through
uniform global pooling to obtain a graph-level representation. However, this
does not explicitly exploit engineering knowledge about characteristic
deformation mechanisms. Bending, torsion, and pumping modes, for example, are
often distinguished by relative responses of roof, floor, side structures, and
longitudinal members rather than by a uniform average over all regions.

To address this limitation, the proposed framework introduces an
\textbf{engineering-aware feature fusion} strategy. In addition to the learned
graph embedding, analytical descriptors are computed directly from the
canonical engineering regions. These descriptors summarise deformation
characteristics used during manual modal assessment, including floor and roof
energy fractions, longitudinal member responses, and vertical deformation
uniformity. Because they are computed at the region level, these descriptors
remain invariant to mesh density and node numbering.

Let $\mathbf{z}_{\mathrm{GNN}}$ denote the global graph embedding obtained by mean pooling over all engineering
region embeddings at the final graph attention layer $L$, and let
$\mathbf{z}_{\mathrm{eng}}$ denote the vector of engineering-informed regional
descriptors. The final graph representation is formed by concatenation,

\[
\mathbf{z}
=
\mathrm{Concat}
\!\left(
\mathbf{z}_{\mathrm{GNN}},\,
\mathbf{z}_{\mathrm{eng}}
\right),
\]

combining learned structural interactions with explicit engineering priors.
This fusion is particularly useful for distinguishing physically similar mode
families where purely learned graph embeddings may not capture subtle but
important engineering differences.

The fused representation $\mathbf{z}$ is passed to a hierarchical prediction
network that follows the reasoning process used during NVH engineering review.
The classifier first predicts the dominant structural mechanism at Level~1,
such as torsion, bending, or pumping, and then refines the prediction to the
corresponding Level~2 subtype. This hierarchical formulation improves
robustness under limited labelled data and reflects the way engineers
typically perform modal assessment.

The complete network is trained end-to-end using a multi-task objective,
$\mathcal{L} = \mathcal{L}_{\mathrm{L1}} + \lambda\,\mathcal{L}_{\mathrm{L2}}$
with $\lambda > 0$, where $\mathcal{L}_{\mathrm{L1}}$ and
$\mathcal{L}_{\mathrm{L2}}$ denote cross-entropy losses for Level~1 and
Level~2 classification respectively, and $\lambda$ is a weighting
hyperparameter balancing the two tasks. The engineering-aware feature fusion
ensures that region-level descriptors provide complementary physical priors
that remain consistent across different mesh discretisations and vehicle
layouts.

\subsection{Engineering Interpretability and Transferability}

A key objective of the proposed framework is not only accurate mode shape
classification, but also the production of representations that remain
meaningful throughout the engineering review process. Because all learning is
performed on the Canonical Engineering Graph Representation, every prediction
can be interpreted, transferred, and reused within the same semantic
engineering representation. The following paragraphs discuss these engineering
properties in turn.

\paragraph{Interpretability.}
Every node in the canonical graph $G$ corresponds to a persistent engineering
region rather than an individual finite element node. Consequently, learned
embeddings, attention coefficients $\alpha_{ij}^{(l)}$, engineering
descriptors, and regional feature importance can be related directly to
recognisable structural components such as roof rails, pillars, floor
structures, side sills, and rear body members. This enables engineers to
verify whether a predicted mode class is supported by physically meaningful
deformation mechanisms, providing a transparent basis for engineering
validation beyond classification accuracy.

\paragraph{Transferability.}
Conventional graph learning approaches typically depend on vehicle-specific
mesh topology, node numbering, or sensor placement. The proposed framework
instead performs learning on the canonical graph that abstracts away from numerical
discretisation. As long as different BiW models can be mapped through
$\Phi$ to the same semantic regional decomposition, the learned
representation remains applicable despite differences in mesh density,
finite element topology, or experimental measurement layout.
Transferability therefore depends primarily on preserving engineering
semantics rather than maintaining identical numerical discretisations.

\paragraph{Knowledge reuse.}
The fused representation $\mathbf{z}$ combines data-driven graph learning with
structural features that engineers already use during modal assessment.
Rather than replacing engineering knowledge, this hybrid representation
preserves physical consistency while allowing the model to accumulate
structural understanding across vehicle programs. Because different vehicle
programs share the same canonical representation, engineering knowledge
learned from one program can be reused and progressively refined for
subsequent programs rather than being relearned from scratch.

\paragraph{Industrial deployment.}
From an industrial perspective, the proposed framework establishes a reusable
engineering representation that can be integrated throughout the engineering
development workflow rather than acting as a task-specific classifier tied to
a single vehicle model. Once a BiW model has been mapped through
$\Phi$, the same canonical graph provides a reusable foundation for
different vehicle programs, simulation variants, experimental measurements,
and related engineering learning tasks without redesigning the underlying
representation.

Although the present work focuses on hierarchical BiW mode shape
classification, the underlying design philosophy is more general.
Engineering problems characterised by persistent engineering entities and
physically meaningful relationships can be represented using a similar
canonical mapping. The proposed framework therefore provides a practical step
towards reusable, explainable, and transferable graph learning for industrial
engineering AI.

%% file: datageneration.tex

\section{Dataset Generation and labeling}

The proposed framework operates on the Canonical Engineering Graph
Representation introduced in Section 3 rather than directly on detailed FE meshes. 
Consequently, the objective of the dataset preparation
process is not only to generate labelled vibration modes, but also to
construct engineering-consistent structural representations that can be mapped
onto the same canonical graph across heterogeneous vehicle programs.

\begin{figure}[t]
  \centering
  \includegraphics[width=\linewidth]{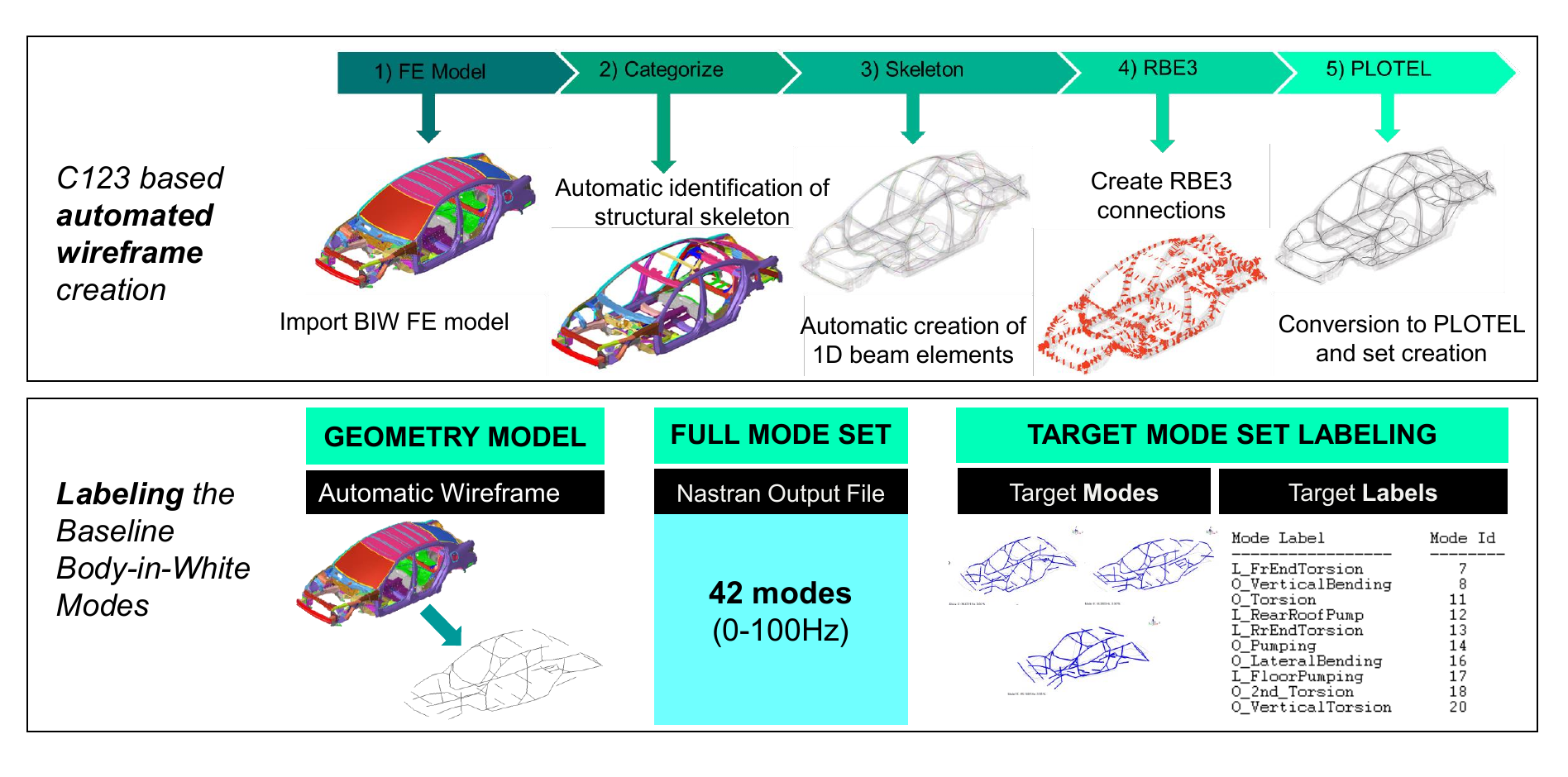}
  \caption{Overview of the engineering wireframe generation, physics-guided
  data augmentation, automatic mode shape labeling, and graph construction
  pipeline.}
  \label{fig:uc1_data}
\end{figure}

The complete data generation pipeline consists of three stages:
engineering wireframe generation from FE models and experimental
measurements, physics-guided data augmentation with automatic label
transfer, and construction of a heterogeneous multi-vehicle benchmark comprising FE 
models and experimental measurements from four vehicle programs.
Figure~\ref{fig:uc1_data} provides an overview of the complete workflow.

\subsection{Automated Wireframe Creation}

Rather than constructing graphs directly from detailed FE meshes, the proposed
framework first extracts an engineering wireframe representation for each
BiW model. The objective is not simply to reduce model
complexity, but to construct a structural abstraction that better reflects the
engineering load paths used during modal assessment while remaining compatible
with both simulation and experimental measurements.

The wireframe represents the principal load-carrying members of
the vehicle, including roof rails, pillars, side sills, floor structures,
cross members, longitudinal rails, and rear body members. Local panels,
brackets, and secondary sheet-metal components are intentionally excluded,
since their dominant contribution is limited to localized deformation and
provides relatively little information for global mode shape classification.

Starting from the imported FE shell model, the structural skeleton is
identified automatically and one-dimensional beam elements are generated along
the centre lines of the principal body members. RBE3 interpolation elements
are subsequently introduced to connect the beam elements to the surrounding
shell mesh without introducing artificial structural stiffness. Finally, the
beam network is converted into PLOTEL elements and organized into engineering
sets, producing a lightweight structural wireframe that remains consistent
across different vehicle variants and assembly configurations.

The wireframe serves as the intermediate structural representation
from which the Canonical Engineering Graph Representation described in
Section 3 is constructed.

\subsection{Physics-guided Data Augmentation and Automatic labeling}

Expert-reviewed mode shape labels are typically scarce in industrial NVH
development because modal interpretation requires experienced CAE engineers.
To alleviate this limitation, a physics-guided concept modelling strategy is
adopted to expand the available training data while preserving physically
meaningful structural behaviour.

The augmentation procedure is based on a reduced modal representation of the
vehicle body, providing an accurate yet computationally efficient description
of structural dynamics in the modal domain
\cite{hadjit2005concept,mucchi2006negative}. Controlled stiffness
modifications are introduced through concept beam models generated using
commercial engineering software. By systematically varying these structural
modifications, physically consistent mode shape variants are created without
requiring repeated full-scale FE analyses. Starting from only 10 expert-reviewed modes, the proposed augmentation
procedure generates 30 structural variants per mode, resulting in
310 labelled mode shapes available for model training. 

A key advantage of this augmentation strategy is that the underlying
engineering wireframe remains unchanged throughout the generation process.
Consequently, one-to-one correspondence between structural entities is
preserved, allowing expert-reviewed labels to be transferred automatically
using the Modal Assurance Criterion (MAC). This avoids repeated manual review
of every generated mode while enabling efficient construction of a
substantially larger labelled dataset.

It should be emphasized that MAC-based label transfer is only applicable
within this controlled augmentation framework, where structural correspondence
is explicitly maintained. Across different vehicle programs, changes in mesh
topology, vehicle geometry, and experimental measurement layouts prevent
direct MAC-based matching. This limitation constitutes one of the primary
motivations for introducing the Canonical Engineering Graph Representation
proposed in this work. 

Mode labeling is restricted to the frequency range of 0--100~Hz, which
contains the dominant global BiW vibration modes considered in the present
study. The selected frequency interval can be adapted for other engineering
applications without changing the underlying graph representation.

\begin{table}[t]
\centering
\caption{Summary of the heterogeneous multi-vehicle dataset. OEM1 node and
label counts are reported before augmentation; following the
physics-guided augmentation procedure, OEM1 contributes 310 labelled
mode shapes for model training.}
\label{tab:dataset_summary}
\small
\begin{tabular}{lcccc}
\toprule
Vehicle & Source & Nodes & Labelled Modes  \\
\midrule
OEM1 & Simulation & 119 & 10  \\
OEM2 & Simulation & 194 & 9  \\
OEM3 & Testing & 58 & 5  \\
OEM4 & Testing & 61 & 2 \\
\bottomrule
\end{tabular}
\end{table}

\subsection{Multi-Vehicle Benchmark Dataset}

The final benchmark comprises four heterogeneous BiW vehicle programs,
including two finite element models and two experimental datasets.
Although these vehicle programs differ substantially in body
geometry, structural topology, node density, and measurement configuration,
they can all be transformed into the same Canonical Engineering Graph
Representation prior to learning.

Table~\ref{tab:dataset_summary} summarizes the characteristics of the
multi-vehicle dataset. The reference vehicle (OEM1) contributes 310 labelled
mode shapes, including the physics-guided augmented samples described above.
Three additional vehicle programs contribute 9, 2, and 5 manually reviewed
mode shapes, respectively. This distribution reflects typical industrial
development, where mature vehicle programs possess extensive historical
annotations, whereas newly introduced vehicle platforms initially contain only
a small number of expert-reviewed vibration modes.

\begin{figure}[t]
\centering
\includegraphics[width=0.57\linewidth]{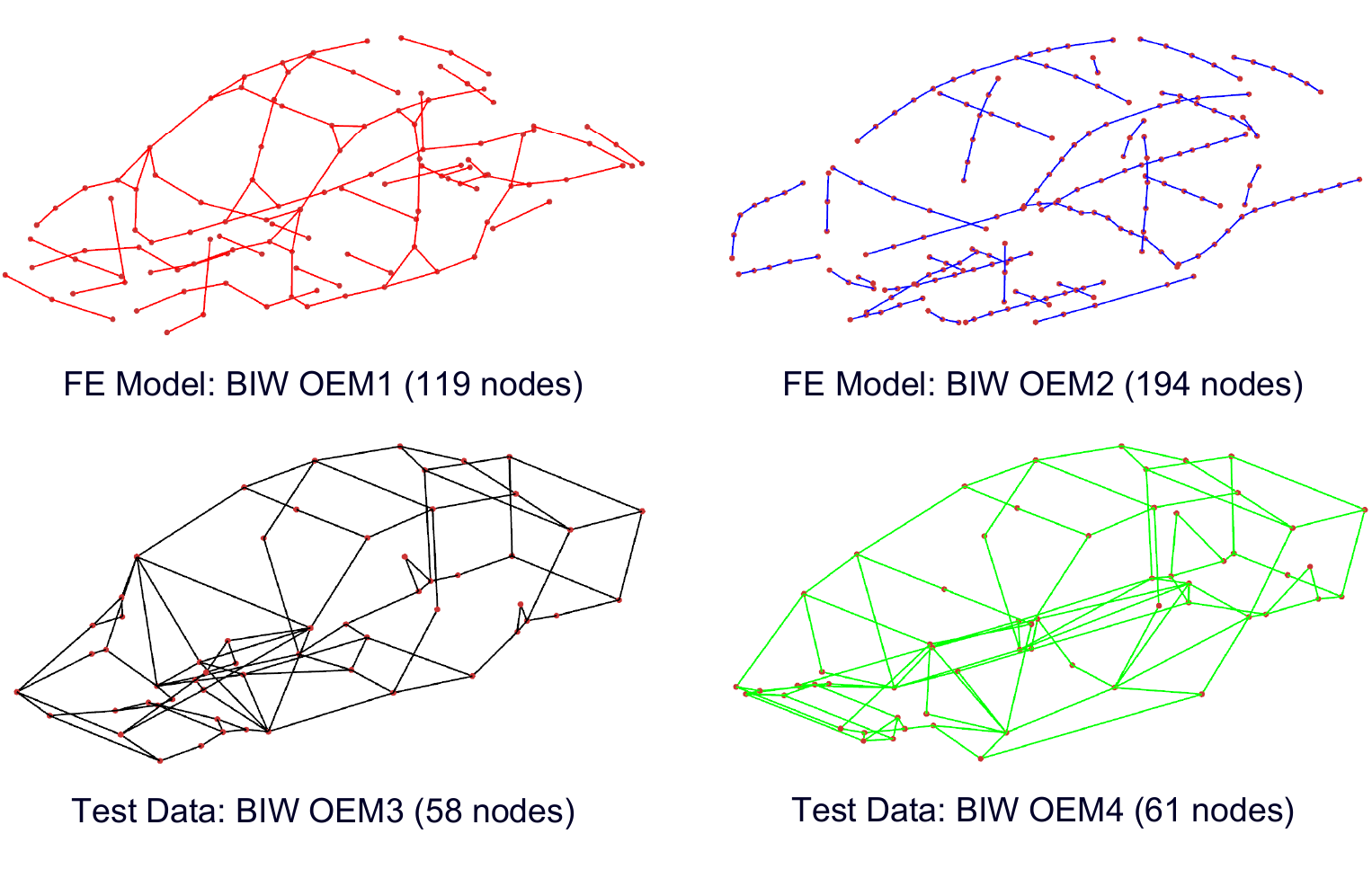}
\hfill
\includegraphics[width=0.4\linewidth]{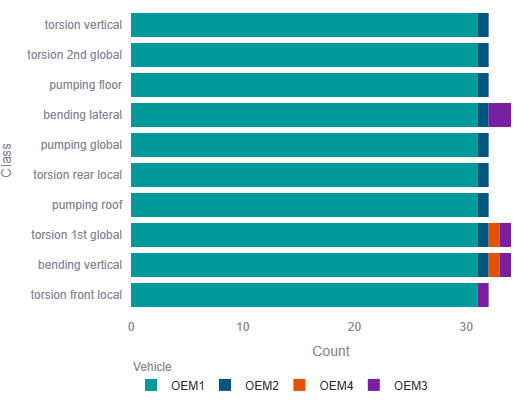}
\caption{Characteristics of the proposed benchmark dataset. (a) Engineering wireframes extracted 
from the four vehicle programs used in this study, illustrating differences in geometry, 
topology, and measurement density across finite element and experimental datasets. 
(b) Distribution of manually reviewed
mode labels across vehicle programs and vibration mode classes.}
\label{fig:dataset_statistics}
\end{figure}

Figure~\ref{fig:dataset_statistics}(a)
illustrates the variation in graph size across vehicle programs, while
Figure~\ref{fig:dataset_statistics}(b) shows the distribution of labelled
samples across vehicle platforms and mode classes. The benchmark therefore
captures three challenges frequently encountered in industrial engineering AI:
heterogeneous engineering representations, limited labelled data, and mixed
simulation-to-test learning.

Training and evaluation follow a stratified split into training,
validation, and test subsets while preserving vehicle diversity
throughout the dataset. More importantly, all vehicle programs are transformed into the same
Canonical Engineering Graph Representation before learning. Consequently, the
subsequent experiments evaluate not only classification accuracy, but also the
ability of the proposed engineering representation to transfer structural
knowledge across heterogeneous finite element models and experimental
measurements under realistic industrial conditions.

%% file: experiment.tex

\section{Experimental Results}

The experimental evaluation aims to validate the three engineering objectives
introduced in Section~3. Specifically, the experiments investigate whether the
proposed Canonical Engineering Graph Representation enables

\begin{enumerate}
\setlength{\itemsep}{0pt}

\item accurate hierarchical mode shape classification under realistic
industrial data constraints;

\item physically meaningful engineering interpretation of the learned
representations; and

\item transferable learning across heterogeneous vehicle programs,
finite element models, and experimental measurements.

\end{enumerate}

All vehicle programs are first transformed into the same Canonical Engineering
Graph Representation before graph learning. Unless otherwise stated, the graph
encoder, training strategy, and region-aware pooling follow the methodology
described in Section~3.

\subsection{Quantitative Evaluation}

Table~\ref{tab:uc1_test_compare} summarizes the quantitative evaluation under
two training strategies. A model trained exclusively on the reference vehicle
achieves high within-vehicle performance but generalizes poorly to unseen
vehicle programs (90.8\% Level-1 and 81.6\% Level-2 accuracy) because of differences in geometry, mesh discretization, and
structural layouts. This observation highlights the limitations of learning
directly from vehicle-specific engineering representations.

In contrast, when all vehicle programs are transformed into the proposed
Canonical Engineering Graph Representation, the multi-vehicle framework
achieves 100\% Level-1 accuracy, 98.7\% Level-2 accuracy, and 99.2\%
combined hierarchical accuracy on the held-out test set (Table~\ref{tab:uc1_test_compare}). Only one of the
76 test samples is misclassified, and the prediction remains within the same
higher-level pumping family, indicating that the learned representation
preserves the dominant structural deformation mechanism even for ambiguous
mode subtypes.

\begin{figure}[t]
  \centering
  \includegraphics[width=0.8\linewidth]{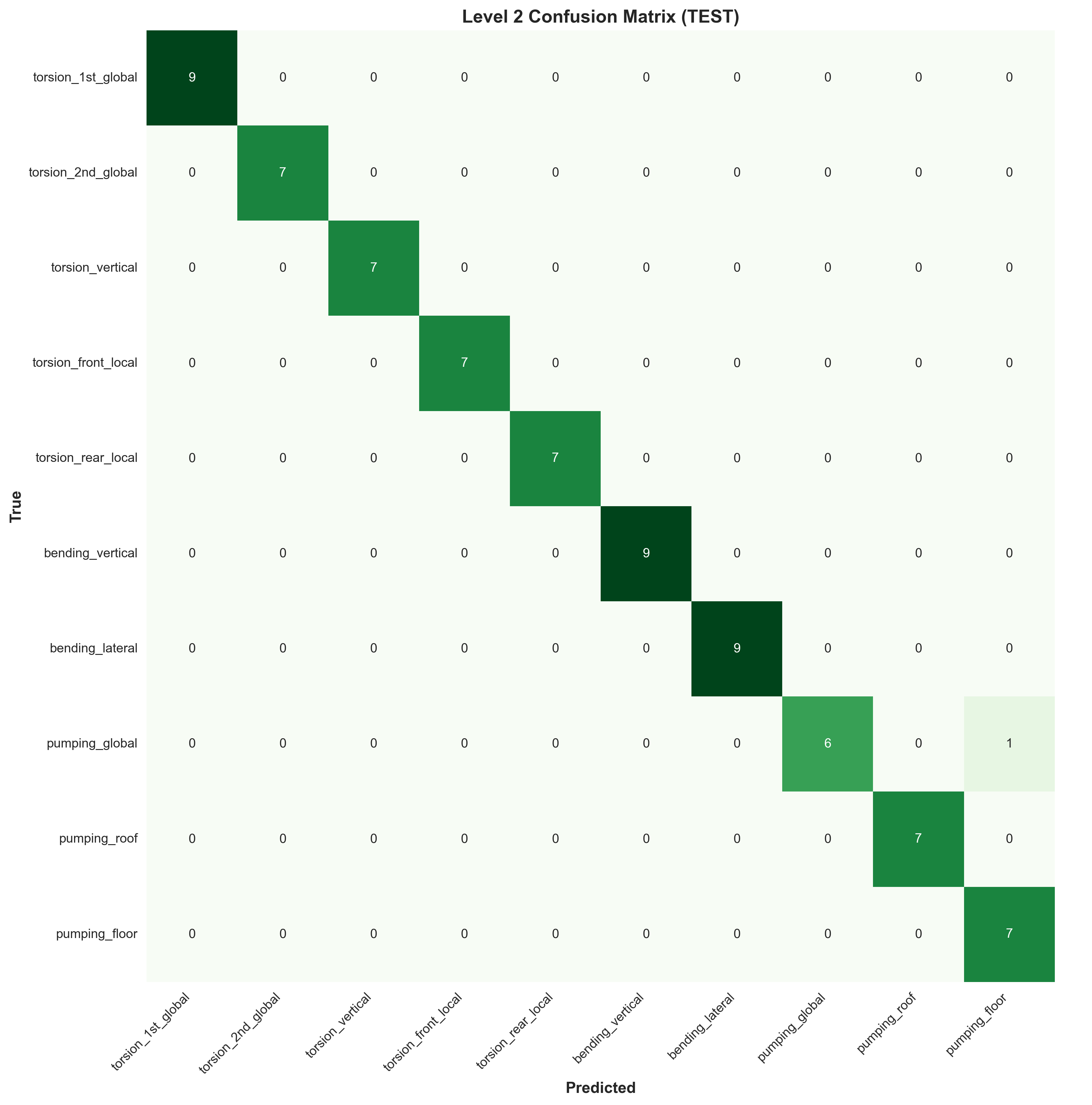}
  \caption{Confusion matrix for fine-grained mode classification on
  the held-out multi-vehicle test set.}
  \label{fig:uc1_test_results}
\end{figure}

The confusion matrix in
Figure~\ref{fig:uc1_test_results} further shows that nearly all errors occur
between physically similar mode families, while torsional and bending modes are
consistently separated. Compared with reference-vehicle-only training, the
proposed framework improves Level-2 classification accuracy by 17.1 percentage
points.

More importantly, this improvement cannot be explained solely by the graph
neural network architecture. Instead, it results from the combination of the
Canonical Engineering Graph Representation, engineering-informed regional
descriptors, and multi-vehicle representation learning. These results support
the central hypothesis of this work that transferable engineering AI depends
primarily on engineering representations rather than geometry-dependent mesh
discretizations.

\begin{table}[h]
  \centering
  \caption{Held-out test performance under different training strategies.}
  \label{tab:uc1_test_compare}
  \scriptsize
  \setlength{\tabcolsep}{3pt}
  \begin{tabular}{lccc}
    \toprule
    Training strategy & L1 Acc. (\%) & L2 Acc. (\%) & Comb. (\%) \\
    \midrule
    Reference-vehicle-only training & 90.8 & 81.6 & 85.3 \\
    Multi-vehicle training & \textbf{100.0} & \textbf{98.7} & \textbf{99.2} \\
    \bottomrule
  \end{tabular}
\end{table}

\subsection{Engineering Interpretability}

Accurate prediction alone is insufficient for practical deployment in
automotive NVH engineering. Engineers must also understand whether model
predictions are supported by physically meaningful structural behaviour.
Unlike conventional graph learning approaches operating directly on finite
element nodes, the proposed framework performs explanation on the Canonical
Engineering Graph Representation, allowing attribution to be visualized
directly on recognizable engineering regions.

\begin{figure}[t]
  \centering
  \includegraphics[width=0.49\linewidth]{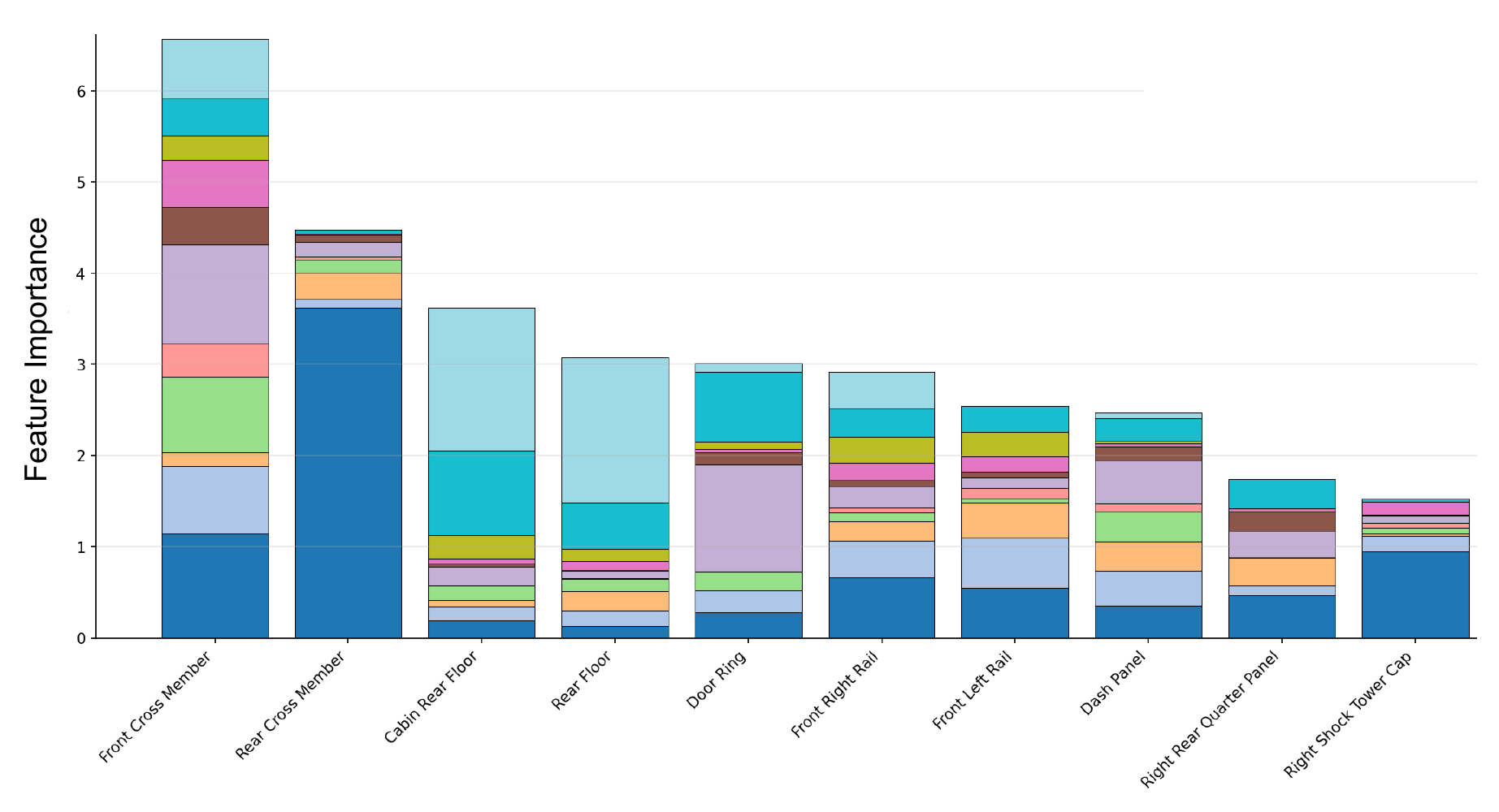}\hfill
  \includegraphics[width=0.49\linewidth]{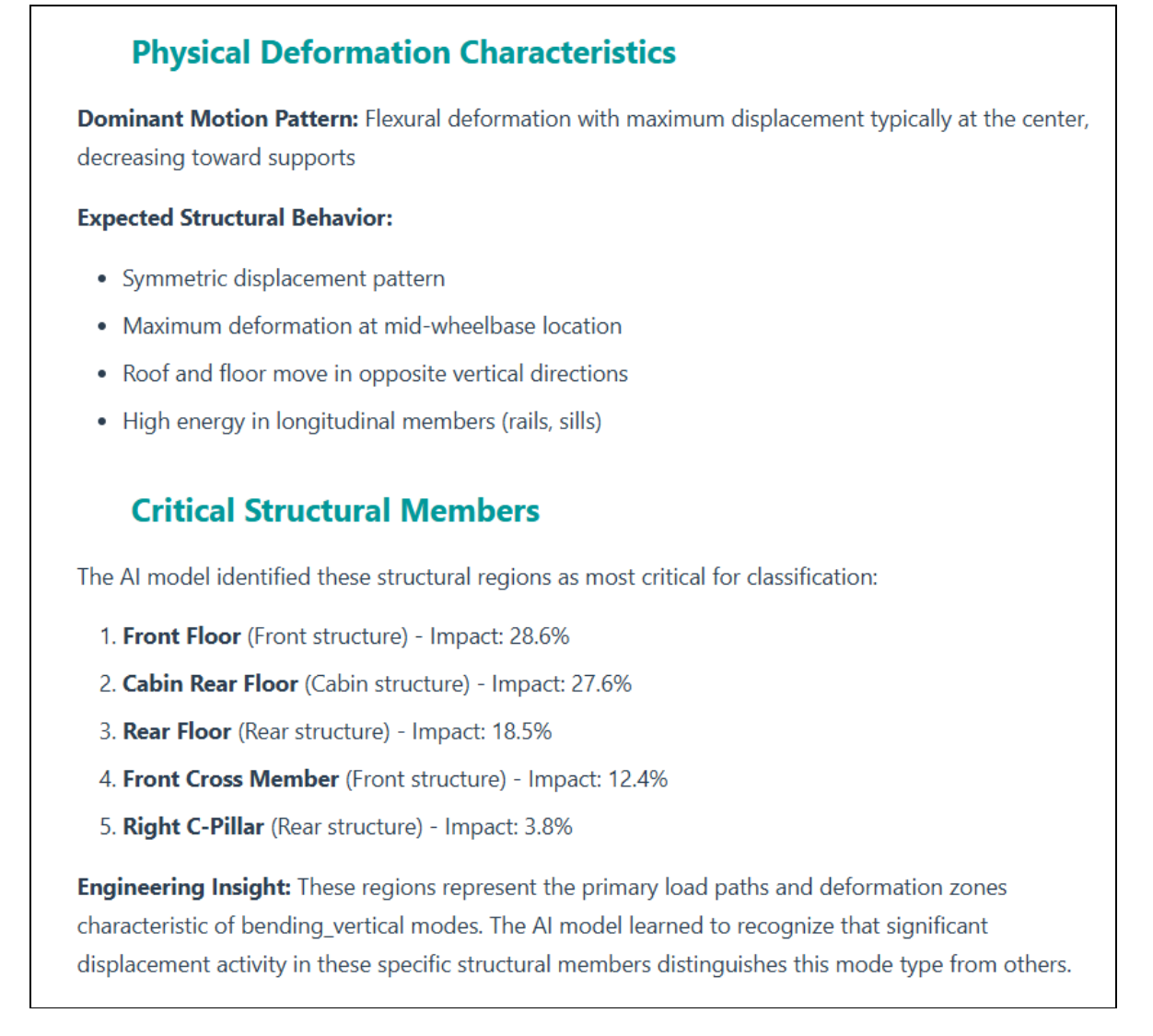}
  \caption{Representative attribution maps illustrating how predictions are
  associated with physically meaningful engineering regions.}
  \label{fig:uc1_explain}
\end{figure}

Representative examples are shown in
Figure~\ref{fig:uc1_explain}. Torsional modes primarily activate
pillar-side sill connections and longitudinal load paths, whereas bending
modes emphasize vertically coupled roof and floor structures. Pumping modes
highlight either floor or roof regions depending on the dominant deformation
mechanism. These activation patterns agree well with established engineering
interpretation of BiW modal behaviour, demonstrating that the learned
representation captures meaningful structural concepts rather than
vehicle-specific numerical artefacts.

From an engineering perspective, this capability provides considerably greater
practical value than conventional black-box classifiers. Rather than returning
only a predicted label, the framework identifies the structural regions that
support the decision, allowing engineers to verify whether the prediction is
consistent with expected deformation mechanisms during manual modal review.

\subsection{Cross-Vehicle Generalization and Transferability}

The principal objective of the proposed framework is not merely to maximize
classification accuracy on a single vehicle program, but to enable engineering
knowledge transfer across heterogeneous engineering datasets. This represents a
significantly more challenging problem because each vehicle differs in body
geometry, finite element topology, node density, and experimental measurement
configuration.

As illustrated in Figure~\ref{fig:dataset_statistics}, all four vehicle programs can
be transformed into the same Canonical Engineering Graph Representation despite
their substantial structural differences. Consequently, graph learning is
performed within a common engineering representation rather than on
vehicle-specific numerical models.

The transfer experiments demonstrate that the proposed representation
successfully preserves engineering semantics across both FE models and
experimental measurements. Despite using only 9, 2, and 5 manually reviewed
mode shapes from the three target vehicle programs, the framework effectively
transfers knowledge from the reference vehicle while maintaining high
classification accuracy. This suggests that the learned representation captures
persistent engineering concepts rather than vehicle-specific mesh
characteristics.

From an industrial perspective, this substantially reduces the effort required
to deploy AI models to new vehicle programs. Instead of constructing a large
labelled dataset for every new vehicle, engineers can initialize the framework
using only a small number of representative mode shapes while reusing
engineering knowledge accumulated from previous development programs.

Overall, the experimental results demonstrate that the proposed Canonical
Engineering Graph Representation satisfies the three engineering objectives
introduced in Section~3. It provides accurate hierarchical mode shape
classification under severe label scarcity, maintains engineering
interpretability throughout the learning process, and enables structural
knowledge to transfer across heterogeneous finite element models and
experimental measurements. These findings provide encouraging evidence that
transferable engineering AI depends primarily on engineering representations
rather than on the numerical discretization of individual simulation models.

%% file: conclusion.tex

\section{Conclusion}

This paper presented a Canonical Engineering Graph Representation for
explainable and transferable engineering AI, demonstrated through the
application of hierarchical Body-in-White (BiW) mode shape classification.
Rather than learning directly from geometry-dependent finite element meshes,
the proposed framework transforms heterogeneous simulation and experimental
data into a common engineering representation composed of persistent
engineering regions and physically meaningful structural relationships. By
separating engineering semantics from numerical discretization, the proposed
representation enables graph learning to operate on engineering concepts
rather than vehicle-specific mesh topology.

Experimental evaluation on four heterogeneous vehicle programs demonstrates
that the proposed representation supports accurate hierarchical mode shape
classification despite limited labeled data and substantial differences in
vehicle geometry, finite element topology, and experimental measurement
layouts. The learned representations remain physically interpretable by
mapping model predictions directly to engineering regions commonly used during
NVH assessment, while successful cross-vehicle transfer indicates that the
framework captures persistent structural concepts rather than
vehicle-specific numerical characteristics.

The principal contribution of this work is therefore the proposed
engineering representation rather than the graph neural network
architecture. The results suggest that transferable
engineering AI depends primarily on engineering representations rather than
geometry-dependent numerical models. The Canonical Engineering Graph
Representation provides a common engineering abstraction through which
historical engineering knowledge can be accumulated, reused, and transferred
across successive vehicle programs and heterogeneous engineering datasets.

Although this study focuses on BiW mode shape classification, the proposed
representation is intentionally formulated as a reusable engineering
foundation rather than a task-specific solution. Engineering problems that can
be described through persistent functional regions and physically meaningful
relationships—including structural dynamics, model correlation, simulation-to-
test analysis, and other CAE workflows—may benefit from the same underlying
representation.

Future work will investigate automatic construction of Canonical
Engineering Graph Representations,
self-supervised and few-shot representation learning,
active data generation for reducing expert labeling effort,
and application to additional simulation-driven engineering domains. 
More broadly, Canonical Engineering Graph Representations are envisioned as an
important step towards trustworthy, explainable, and
reusable engineering AI capable of integrating heterogeneous engineering
knowledge throughout the development workflows.

\section*{Acknowledgment}

This work is supported by the Flanders Innovation and Entrepreneurship (VLAIO)
project \emph{Simulation And TestIng Solutions For TrustworthY Data-centric AI
(SATISFY.AI)}.